\documentclass[twocolumn,showpacs,preprintnumbers,amsmath,amssymb]{revtex4}
\usepackage{graphicx}% Include figure files
\usepackage{dcolumn}% Align table columns on decimal point
\usepackage{bm}% bold math
\usepackage[usenames]{color}

\begin{document}

%\preprint{APS/123-QED}

\title{Orbital degrees of freedom as
origin of magnetoelectric coupling in magnetite}
%$R$-dependence of magnetoelectric coupling in orthorhombic $R$MnO$_{3}$ \\ 
%(\textit{R} = La, Pr, Nd, Sm, Gd, Tb, Dy, Ho, Er, Tm, Yb, Lu)}% Force line breaks with \\   \textit{}
%(\textit{R}=Rare-Earth Ions)}

\author{Kunihiko Yamauchi}
\author{Silvia Picozzi}%
%\author{Someone Else$^2$}
% \altaffiliation[]{}%Lines break automatically or can be forced with \\
%\author{Silvia Picozzi}%
% \email{silvia.picozzi@aquila.infn.it}
\affiliation{
Consiglio Nazionale delle Ricerche (CNR-SPIN), 67100 L'Aquila, Italy}
%2. Somewhere
%\author{Anyone Else}
%\affiliation{
%somewhere\\

\date{\today}% It is always \today, today,
\newcommand{\fo}{Fe$_{3}$O$_{4}$ }
\begin{abstract}
%An article usually includes an abstract, a concise summary of the work
%covered at length in the main body of the article. It is used for
%secondary publications and for information retrieval purposes. Valid
%PACS numbers may be entered using the \verb+\pacs{#1}+ command.

%{\em Abstract: Later on we will decide how to put it} \\

%{\bf 
%Along the exciting roadmap leading to a novel generation of electrically-controlled spintronic devices,
%a microscopic understanding of magnetoelectricity, {\em i.e.} 
% the coupling between magnetic (electric) properties and external electric (magnetic) fields,  is a crucial milestone. 
A microscopic understanding of magnetoelectricity, {\em i.e.} the coupling between magnetic (electric) properties and external electric (magnetic) fields,  is a crucial milestone
% in the field of electrically-controlled spintronic devices. 
 for future generations of electrically-controlled spintronic devices. 
Here, we focus on the first magnetoelectric known to mankind: magnetite.
By means of  a joint approach based on phenomenological Landau theory and density-functional simulations, we show that magnetoelectricity in charge-/orbital-ordered Fe$_{3}$O$_{4}$ in the non-centrosymmetric $Cc$ structure is driven by the interplay between a peculiar orbital-order and on-site spin-orbit coupling.
The excellent agreement with available experiments confirms our theoretical picture, pointing to magnetite as a prototype of  
a novel category of magnetoelectrics where ferroelectric polarization can be induced, tuned or switched via a magnetic field. 
%}
\end{abstract}

\pacs{Valid PACS appear here}% PACS, the Physics and Astronomy
\maketitle

%\section{\label{sec:level1}First-level heading:\protect\\ The line
%break was forced \lowercase{via} \textbackslash\textbackslash}
%\section{\label{sec:intro}Introduction\protect\\}
Magnetoelectric (ME) effects --- {\em i.e.} how to control magnetic (electric) properties via electric (magnetic) fields  --- have been intensively investigated in recent years, often in connection
 to {\em multiferroics}, materials where two or more ferroic properties ({\em i.e.} long range spontaneous magnetism, deformation or ferroelectricity) coexist in the same phase.\cite{maxim,silvia,fiebig.ME.review} 
In this framework, it was recently found ferrimagnetic magnetite (Fe$_{3}$O$_{4}$) to show a
sizable ferroelectric (FE) polarization at low temperatures, with the microscopic mechanism clarified theoretically:\cite{Alexe, yamauchi.prb}
the FE polarization, of the order of few $\mu$C/cm$^{2}$, is induced purely by the non-centrosymmetric charge-order (CO), a mechanism that is rarely seen in other materials.\cite{lufeo}

The ME effect in magnetite was discovered by Rado and Ferrari in 1975 \cite{rado.prb1975}; 
however, its importance has been neglected for long time.  It is the aim of this paper to clarify the microscopic origin of this peculiar ME effect in magnetite, by means of Landau theory and density-functional-theory (DFT).  We start from two important observations made in Ref. \onlinecite{rado.prb1975}: 
(i) The electric polarization $\bm P$ behaves as a nonlinear function of the applied magnetic field $\bm H$, more precisely, $P_{a}\propto\sin^{2}\theta$ (with negligible $\sin^{4}\theta$ term), with $\theta$ denoting the polar angle  of  the $\bm H$ direction at 4.2K.
%and exists even in the absence of applied electric field at 4.2K.
(ii) The observed $P$ (reported in arbitrary units) is saturated with respect to a certain $H^{\rm sat}$, {\em i.e.} its behavior being similar to the magnetization $M$.  
This implies that $\bm P$ arises in the perfectly ``collinear'' ferrimagnetic configuration, with all spins simultaneously following the direction of $\bm H$ (i.e. $\bm M$//$\bm H$); this
shows a distinct contrast with respect to many cases of single-phase magnetoelectrics, where either $\bm H$-induced ``non-collinear'' spin canting or spin spiral modulation occur (e.g. at Cr$_{2}$O$_{3}$\cite{mostovoy.cr2o3}, LiNiPO$_{4}$\cite{yamauchi.linipo4} and TbMnO$_{3}$\cite{kimura.tbmno3}). 
%They have associated the ME effect to the single-site magnetic anisotropic energy. 
% the Fe spins are all parallel/anti-parallel coupled and simultaneously following the  $\bm H$ direction when $H$ is enough large 
%although, in most cases, single-domain ME effect occurs with either spin canting with respect to the antiferromagnetic configuration or spin spiral case.  

%
%\begin{figure}[!h]
\begin{figure}[!h]
%\vspace{0.2cm}
%\resizebox{75mm}{!}
%\resizebox{72mm}{!}
\resizebox{70mm}{!}
%\resizebox{75mm}{!}
{
\includegraphics{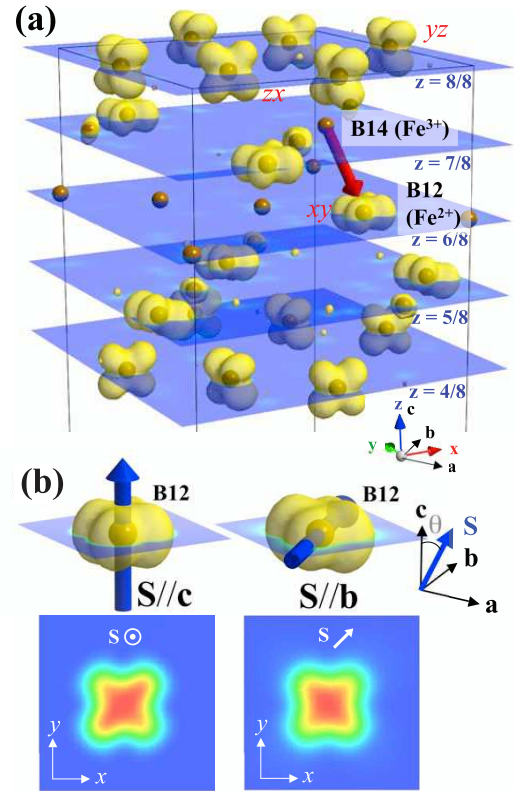}
}
\caption{\label{fig:parchg} 
(a) Charge/orbital ordering shown by an isosurface of charge density of  Fe-$t_{2g}^{\downarrow}$ states in the upper half ($1/2$$<$$z$$<1$) of the primitive $Cc$ unit cell (for details, see ref. \onlinecite{yamauchi.prb}).
%The lower half (not shown) is related to the upper half through the $c$-glide, \{$\sigma_{2b}+(00\frac{1}{2})$\} (for details, see ref.\cite{yamauchi.prb}.) 
The red arrow shows the electric dipole connecting B12(Fe$^{2+}$) and B14(Fe$^{3+}$) sites, responsible for the ferroelectric $\bm P$. Vectors ($\bm a$, $\bm b$, $\bm c$)
% (($\bm x$, $\bm y$, $\bm z$)) 
denote directions of the conventional 
%(primitive)
 lattice vectors, linked to the cartesian coordinate system through 
 %$\bm x$=$\frac{\bm a+\bm b}{2}$,  $\bm y$=$\frac{-\bm a+\bm b}{2}$ and $\bm z\simeq\bm c$. 
 $\bm x$=${\bm a+\bm b}$,  $\bm y$=${-\bm a+\bm b}$ and $\bm z\simeq\bm c$ (due to a small monoclinic distortion). 
The occupied orbital state at some Fe$^{2+}$ sites is labeled as $xy$/$yz$/$zx$, used as site index hereafter. 
(b) Change of the charge density (an isosurface and a section) on Fe-B12 site upon rotation of the spin direction (shown by blue arrows in the upper panel). To emphasize the effect, SOC term is enhanced by a factor of 10. 
%The change in the $d$-occupancy modulates the size of the electric dipole and the consequent $\bm P$, leading to the ME effect. 
}
\vspace{-0.6cm}
\end{figure}

{\em Structural Details.} --- Magnetite shows a well-known metal-insulator transition, namely the ``Verwey transition'' at $T_{\rm V}$=120K, below which the crystal structure changes from cubic $Fd3m$ to a less symmetric structure; correspondingly, the  Fe$^{2+}$/Fe$^{3+}$ charge ordering, observed on Fe-B sites in the inverse-spinel AB$_{2}$O$_{4}$ lattice,\cite{Verwey, Garcia} arises.
Despite several different space groups were proposed for the low-temperature structure,\cite{Garcia} 
%bunch of experimental studies suggesting various possible low-temperature structures\cite{Garcia}, 
 a base-centered monoclinic $Cc$ symmetry (\#9) was suggested as ground state %of lattice structure 
 both by experimental and theoretical studies.
 %\cite{rixs, zuo, Jeng_prl, Jeng_prb} ! Jeng_prl is calculation for P2/c !
 \cite{zuo, Jeng_prb} 
The absence of inversion symmetry in the $Cc$ space group and the consequent polar CO pattern were found to be relevant for the presence of ferroelectricity.\cite{Alexe}  
Indeed, the FE polarization is primarily caused by uncompensated local electric dipoles connecting 
Fe$^{2+}$ and Fe$^{3+}$ ions at B sites; %, which don't have the counterpart with the equivalent charge, connected by space inversion; 
as shown in Fig.\ref{fig:parchg}(a) and deeply discussed in Refs.[\onlinecite{yamauchi.prb, B12B14}], Fe$^{2+}$ at B12 site and Fe$^{3+}$ ions at B14 site form one of the dipoles responsible for the net $\bm P$ in the $Cc$ unit cell. Fig.\ref{fig:parchg}(a) also shows
the orbital order (OO) at Fe$^{2+}$ ($d^{6}$) sites. 
Being coupled to cooperative Jahn-Teller (JT) distortions,  
the partially filled minority-spin $t_{2g}^{1\downarrow}$ orbital shows one of the three $d_{xy}$/$d_{yz}$/$d_{zx}$ charge distributions. 
In the primitive unit cell, 32 Fe-B sites are split into 16 Fe$^{3+}$ sites and 16 Fe$^{2+}$ sites; in addition,
the latter JT-active sites are furthermore split into three groups: 8 ${xy}$, 4 ${yz}$ and 4 ${zx}$ sites.\cite{suffix} 
%
%(a) previously reported,\cite{yamauchi.prb} both the $P2/c$ and $Cc$ structures show CO and orbital ordering (OO) at Fe-B sites. 
%In the unit cell, 32 Fe-B-sites are split into 16 Fe$^{3+}$ ($d^{5}$) sites and 16 Fe$^{2+}$ ($d^{6}$) sites. 
%The latter are Jahn-Teller (JT) active ions, where the partially filled down-spin $t_{2g}^{1\downarrow}$ shell shows one of $d_{xy}$/$d_{yz}$/$d_{zx}$ charge distribution, according to the related JT distortion of the FeO$_{6}$ octahedral cage. Majority spin states, on the other hand, are completely filled. 
%In this context, the JT-acrive 16 Fe$^{2+}$ sites are furthermore split into three groups: 8 with $d_{xy}$, 4 with $d_{yz}$ and 4 with $d_{zx}$ OO. 
The OO pattern
% (in this paper, the suffixes $x$, $y$, $z$ are always consistent with the vectors shown in Fig.\ref{fig:parchg}) 
is such that the orbitals avoid to overlap, consistent with reducing the inter-site Coulomb repulsion and optimizing cooperative JT distortions. 
Consequently, these orbitals do not fully lie in the $xy$/$yz$/$zx$ planes, but are rather slightly tilted.

{\em Macroscopic Model.} --- Hereafter, we separate the spontaneous CO-induced, $\bm P^{\rm CO}$, and the $\bm M$-direction-dependent, $\bm P^{\rm ME}(\bm M)$. 
According to Ref. \onlinecite{jia.nagaosa}, 
the mechanisms leading to $\bm P^{\rm ME}$ are generally classified as driven by magnetostriction ($P^{\rm MS}$), spin-current ($P^{\rm sp}$) and orbital degrees of freedom ($P^{\rm orb}$).  
%As taking into account the observed $\bm P$ saturation with $\bm H$\cite{rado.prb1975}, 
In our case, the collinear ferrimagnetic configuration rules out both $P^{\rm MS}$ and $P^{\rm sp}$, 
whereas only $P^{\rm orb}$ can be induced by the spin-orbit coupling (SOC) within orbital-unquenched $t_{2g}^{4}$$e_{g}^{2}$ state at Fe$^{2+}$($d^{6}$) sites. Summarizing, the total polarization is here described as ${\bm P}^{\rm total}({\bm M})
%={\bm P^{\rm CO}}+{\bm P^{\rm ME}}({\bm H}) 
= {\bm P^{\rm CO}}+{\bm P^{\rm orb}}({\bm M})$. 
%%%The $\bm P^{\rm orb}$ term shows up in YVO$_{3}$ together with $\bm P^{\rm sp}$ term via Dzyaloshinski-Moriya (DM)  interaction\cite{ren.yvo3}, 
%%%whereas it is the first case that sorely $\bm P^{\rm orb}$ gives the ME effect, to the best of our knowledge. %, at magnetite. 
As we will discuss later, the ${\bm P^{\rm orb}}$ term is deeply related to the single-site magnetic anisotropy energy (MAE), both having SOC as common underlying origin. 
\begin{table}[h]
\vspace{-0cm}
\caption{
Matrices of the generators of the $Cc$ space group  in the representations spanned by $M$ and $P$. 
The space group elements are denoted as the identity $E$ and $c$-glide=$\{\sigma_{b}|00\frac{1}{2}\}$, with time reversal $T$. 
\label{table:group}
}
\begin{center}
\begin{tabular}{|c|cccc||c|cccc|}
\hline
	& 	$E$  &$c$ & $TE$ & $Tc$ & &$E$  &$c$ & $TE$ & $Tc$ \\
\hline
$M_{a}$, $M_{c}$ &1&-1&-1&1  &$P_{a}$, $P_{c}$ &1&1&1&1\\
$M_{b}$&1&1&-1&-1	&$P_{b}$&1&-1&1&-1\\
%$M_{c}$&1&-1&-1&1  &$P_{c}$&1&1&1&1\\
\hline
%$P_{a}$&1&1&1&1\\
%$P_{b}$&1&-1&1&-1\\
%$P_{c}$&1&1&1&1\\
%\hline
\end{tabular}
\end{center}
\label{default}
%to save space 
\vspace{-0.6cm}
\end{table}

In order to shed light on the peculiar ME effects in ${\bm P^{\rm orb}}$, we briefly show the group theory analysis 
 and a derivation  based on the Landau theory of phase transitions.\cite{Landau} 
When working in the $Cc$ space group with the symmetry operations \{$E$, $c$\}, 
the ferrimagnetic order 
%(with the spin angle $\theta$) 
leads to a lowered symmetry in the magnetic space group, with the presence of SOC. 
%At the hypothetical transition from the nonmagnetic- to the ferrimagnetic-phase, 
We define the order parameter,  
${\bm M}= -\sum_{i}{\bm S}^{A}_{i} + \sum_{i}{\bm S}^{B}_{i}$, as a linear combination of Fe spins at A and B sites. 
 %  applied to the ferrimagnetic order (with the spin angle $\theta$) and the coupling to the electric polarization $\bf P$ in magnetite. 
%We don't take into account the spontaneous ferroelectric polarization due to the charge-order, which exists irrespectively of the spin direction. 
%We work in the $Cc$ (\#9) space group setting, with the symmetric operations, \{$E$, $c$\}. 
%, where $c$ denotes the $c$-glide, $\{\sigma_{y}|00\frac{1}{2}\}$. 
%Here, the hypothetical transition from nonmagnetic phase to ferrimagnetic phase leads to lowered symmetry in magnetic space group. 
%The ferrimagnetic order parameter of $\bf M$ is written as a linear combination of Fe spins ${\bf S}_{i}^{A}$ and ${\bf S}_{i}^{B}$ located on Fe-A sites and Fe-B sites, respectively, in the unit cell, 
%\begin{equation}
%{\bf M} = -\sum_{i}^{16} {\bf S}^{A}_{i} + \sum_{i}^{32} {\bf S}^{B}_{i}. 
%\end{equation}
%The  components of $\bf M$ and $\bf P$ transform according to the rules given in Table \ref{table:group}. 
Using the transformation rules  given in Table \ref{table:group}, we analyze, in the thermodynamic free energy, the possible ME coupling terms of the form
${\bm P} \cdot {\bm M}^{2}$, 
%\color{red} 
which are invariant under symmetry operations.
%\color{black: 
%
%
%\begin{equation}% \vspace{-1cm}
%\begin{split}\label{eq:Fme1}
\vspace{-0.5cm}
\begin{align}\label{eq:Fme1}
%F_{\rm ME} &=&c_{aa}P_{a}M_{a}^{2} + c_{ab}P_{a}M_{b}^{2} + c_{ac}P_{a}M_{c}^{2}\notag \\
%&+& c_{aac}P_{a}M_{a}M_{c}\notag \\
%&+& c_{bab}P_{b}M_{a}M_{b} + c_{bbc}P_{b}M_{b}M_{c} \notag \\
%&+& c_{ca}P_{c}M_{a}^{2} + c_{cb}P_{c}M_{b}^{2} + c_{cc}P_{c}M_{c}^{2} \notag \\
%&+& c_{cac}P_{c}M_{a}M_{c}, 
%
%F_{\rm ME} &=&c_{aa}P_{a}M_{a}^{2} + c_{ab}P_{a}M_{b}^{2} + c_{ac}P_{a}M_{c}^{2}\notag \\
%&+& c_{ca}P_{c}M_{a}^{2} + c_{cb}P_{c}M_{b}^{2} + c_{cc}P_{c}M_{c}^{2} \notag \\
%&+& c_{aac}P_{a}M_{a}M_{c}+ c_{cac}P_{c}M_{a}M_{c}\notag \\
%&+& c_{bab}P_{b}M_{a}M_{b} + c_{bbc}P_{b}M_{b}M_{c}, 
%
F_{\rm ME}&= c_{aa}P_{a}M_{a}^{2} + c_{ab}P_{a}M_{b}^{2} + c_{ac}P_{a}M_{c}^{2}+ c_{ca}P_{c}M_{a}^{2} \notag\\
+& c_{cb}P_{c}M_{b}^{2} + c_{cc}P_{c}M_{c}^{2} + c_{aac}P_{a}M_{a}M_{c} \notag \\
+& c_{cac}P_{c}M_{a}M_{c}+ c_{bab}P_{b}M_{a}M_{b} + c_{bbc}P_{b}M_{b}M_{c}, 
%\end{split}
%\end{equation}
\vspace{-0.5cm}
\end{align}
whereas the dielectric energy is traditionally written as:
%\begin{eqnarray}\label{eq:Fde}
$F_{\rm DE} =- {\bm P}^{2}/2\chi$,  
%\end{eqnarray}
where the $c_{ij}$, $c_{ijk}$ coefficients and $\chi$ 
%\color{red}
(set as 1 in what follows) 
%\color{black}
are constants. 
The minimum of $F$=$F_{\rm ME}$+$F_{\rm DE}$ occurs  when
 $\partial F/\partial P_{a}=\partial F/\partial P_{b}=\partial F/\partial P_{c}=0$, so that $\bm P$ is obtained. 
% , as leading to
%\begin{eqnarray}\label{eq:P}
%P_{a}&=&c_{aa}M_{a}^{2}+c_{ab}M_{b}^{2}+c_{ac}M_{c}^{2}+c_{aac}M_{a}M_{c}, \notag \\
%P_{b}&=&c_{bab}M_{a}M_{b}+c_{bbc}M_{b}M_{c}, \notag \\
%P_{c}&=&c_{ca}M_{a}^{2}+c_{cb}M_{b}^{2}+c_{cc}M_{c}^{2}+c_{cac}M_{a}M_{c}. 
%\end{eqnarray}
%where the term dependent on ${\bf M}^{2}$, which is not relevant to the spin angle,  is neglected. 
After assuming a simultaneous rotation of Fe spins in the $bc$ plane by an angle $\theta$ with respect to the $c$ axis,
%keeping the ferrimagnetic order, 
setting
%\begin{equation}
%-{\bf S}^{A}/S^{A}={\bf S}^{B}/S^{B}=(0, \sin\theta, \cos\theta). 
%\end{equation}
%When we describe the net magnetization as 
%$M=-16S^{A}+\sum_{i}^{32} {S}^{B}_{i}$ (note that the size of the spins at Fe-B sites varies,  due to CO), ${\bf M}$ is also described as 
%\begin{equation}
${\bm M}=M(0, \sin\theta, \cos\theta)$,  
%\end{equation}
%Then, Eq. (\ref{eq:P}) becomes 
we derive
%\color{red}
\begin{align}\label{eq:P2}
\vspace{-0.5cm}
%P_{a}&=&M^{2} ( c_{ab}\sin^{2}\theta+c_{ac}\cos^{2}\theta ), \notag \\
P_{a}(\theta)&=\tfrac{M^{2}}{2}  ( -c_{ab}+c_{ac})\cos2\theta +\tfrac{M^{2}}{2}(c_{ab}+c_{ac}), \notag \\
%P_{y}&=&M^{2}  c_{yyz}\sin\theta\cos\theta =1/2M^{2}  c_{yyz}\sin2\theta, \notag \\
P_{b}(\theta)&=\tfrac{M^{2}}{2}  c_{bbc}\sin2\theta, \notag \\
%P_{c}&=&M^{2} ( c_{cb}\sin^{2}\theta+c_{cc}\cos^{2}\theta ).
P_{c}(\theta)&=\tfrac{M^{2}}{2}  ( -c_{cb}+c_{cc})\cos2\theta +\tfrac{M^{2}}{2}(c_{cb}+c_{cc}). 
\vspace{-0.5cm}
\end{align}
%
%\color{black}
%\\
%If we assume the special case, $c_{ab}=-c_{ac}$ and $c_{cb}=-c_{cc}$, Eq.(\ref{eq:P2}) can be reduced to the simple form as;
%\begin{eqnarray}\label{eq:P3}
%P_{a}&=&-2M^{2} c_{ab}\cos2\theta , \notag \\
%%P_{y}&=&M^{2}  c_{yyz}\sin\theta\cos\theta =1/2M^{2}  c_{yyz}\sin2\theta, \notag \\
%P_{b}&=&M^{2}/2  c_{bbc}\sin2\theta, \notag \\
%P_{c}&=&-2M^{2} c_{cb}\cos2\theta, 
%\end{eqnarray}
Remarkably, the $\theta$-dependence perfectly agrees with the experimentally observed property, $P_{a}\propto\sin^{2}\theta$.\cite{rado.prb1975}
% our DFT results, as shown in the main paper. 
Incidentally, we note that other ME systems often need two or more antiferromagnetic order parameters 
%, e.g., ${\bf E}_{1}={\bf S}_{1}-{\bf S}_{2}-{\bf S}_{2}+{\bf S}_{4}...$ and ${\bf E}_{2}={\bf S}_{1}+{\bf S}_{2}-{\bf S}_{2}-{\bf S}_{4}...$  (with components belonging to different representations) are important 
to cause ME effects\cite{silvia.prl}, whereas here only one magnetic order parameter $\bm M$ is relevant.

Now, selecting the $\theta$-dependent terms
% related to the ME effect 
from Eq.(\ref{eq:Fme1}), we obtain;
\begin{align}\label{eq:Fme2}
%F_{\rm ME} = M^{2}/2\{( - c_{xy}P_{x}  + c_{xz}P_{x} - c_{zy}P_{z} + c_{zz}P_{z})\cos2\theta \notag\\
% + c_{yyz}P_{y}\sin2\theta\} \notag \\
% &F_{\rm ME}(\theta) = \tfrac{M^{2}}{2}(a_{\bf P}\cos2\theta  + b_{\bf P}\sin2\theta),  
  &F_{\rm ME}(\theta) = \tfrac{M^{4}}{2}(a_{\bf P}\cos2\theta  + b_{\bf P}\sin2\theta),  
 \end{align}
 %; \notag \\
%\end{eqnarray}
%with
%\begin{eqnarray} 
where $ a_{\bf P}=(- c_{ab}  + c_{ac})P_{a} + ( - c_{cb} + c_{cc})P_{c}$ and 
$b_{\bf P}=c_{bbc}P_{b}.$ 
%\end{align}
Here, the $\theta$-dependence of $F_{\rm ME}$ is nothing but the MAE, i.e., 
we obtained an expression for the MAE as a function of $\bm P$. 
%Therefore, the result obtained here by assuming a $\bf P \cdot M^{2}$ form in the thermodynamic free energy 
%is consistent 
This is consistent with what already proposed by Rado {\em et. al.}, where they assumed the electric field $\bm E$-dependence of the MAE coefficient $K'_{b}$.\cite{rado.prb1975}  
Our additional advantage here is that we obtain the  $\theta$-dependence of $\bm P$ starting from the polar $Cc$ symmetry, which was unknown at that time. 
Also note that none of the experimentally proposed centrosymmetric structures (e.g., $P2/c$ or $Pnma$) allows for  $\bm M$-induced $\bm P$, because the inversion symmetry is not broken by the ferrimagnetic order ({\rm all Fe-B spins are parallel}). Our findings related to ME effects are therefore an indirect confirmation that, among those that were experimentally put forward, the $Cc$ symmetry is the ground state of the CO-phase.
%The difference between the two studies is that we here derived the $\theta$-dependence of ($P_{a}$, $P_{b}$, $P_{c}$) by taking into account the symmetry of the $Cc$ space group. 
% 

{\em DFT calculations} ---
In order to quantitatively evaluate the $\bm P(\theta)$ behavior and to investigate the microscopic mechanism, DFT calculations were performed using the VASP\cite{vasp} code. 
%with projector-augmented-wave (PAW) basis set and the Perdew-Becke-Erzenhof (PBE) potential\cite{pbe}. 
%For the generalized gradient approximation 
%with GGA+$U$ approach\cite{ldau}. 
Starting from a previous study by Jeng {\em et al.}\cite{Jeng_prb}, 
we used their optimized $Cc$ structural parameters and the GGA+$U$\cite{ldau} approach, with $U=4.5$ eV and $J=0.89$ eV for  Fe-$d$ state (Fe-$3d^{6}4s^{2}$ electrons are treated as valence). The
2$\times$2$\times$1 Monkhorst-Pack $k$-point grid was used; other details are the same as in Ref. \onlinecite{yamauchi.prb}. 
%the parameters $U=4.5$ eV and $J=0.89$ eV were taken from Ref.\cite{Jeng_prl}. 
%To calculate the FE polarization, we compared the Berry phase\cite{berry} between paraelectric (PE) ($P2/c$) and FE  ($Cc$) phases in the primitive cell of the base-centered monoclinic $Cc$ lattice 
%(with 112 atoms/cell, containing 16 Fe$^{3+}$ ions on A sites, 16 Fe$^{3+}$ ions and 16 Fe$^{2+}$ ions on B sites). 
%
%A cutoff energy of 400 eV for plane waves and a 2$\times$2$\times$1 Monkhorst-Pack $k$-point grid were used.  
%Fe-$3d^{6}4s^{2}$ and O-$2s^{2}2p^{4}$ electrons were treated as valence: 
%For a better convergence of the charge density, semi-core Fe-$3p^{6}$ electrons were treated as core electrons,
% whereas these same electrons were treated as valence in our previous study.\cite{yamauchi.prb} 
%As using the core-treatment of Fe-$3p^{6}$ and the same computational parameters, 
%The resulting properties (structure and orbital shape, etc) are almost identical to Ref.\cite{Jeng_prl} by Jeng {\em et al.}, therefore we used their optimized $Cc$ structural parameters ($i.e.$ lattice parameters, 
% $a$=11.8838\AA, $b$=11.8682\AA, $c$=16.7369\AA\ and $\beta$=90.2006$^{\circ}$, and the internal coordinates). 
%Incidentally, 
%to be consistent with old experimental references, where they have defined $a_{m}$, $b_{m}$ and $c_{m}$ axis as magnetically hard, intermediate and easy axis, respectively, without knowing information of the low-temperature crystal structure, 
%we should interchange $a$ and $b$ of our $Cc$ structure setting; $a_{m}$=$b$, $b_{m}$=$a$ as we know the magnetic anisotropy later. 
A collinear ferrimagnetic configuration was set, with all Fe-B  (Fe-A) sites as up-spin (down-spin) sites. 
By introducing SOC self-consistently, %all spins were rotated simultaneously, 
%by constraining penalty energy, 
%so as to 
we evaluated both {\em i} ) the MAE through the total-energy change and {\em ii}) the ME effect through the change of $\bm P$ calculated by the Berry phase method, \cite{berry}
upon rotation of the $\bm M$ direction with respect to the crystalline axes. 

\begin{figure}
\resizebox{82mm}{!}
%{\includegraphics{figure/DFTall1.eps}}
{\includegraphics{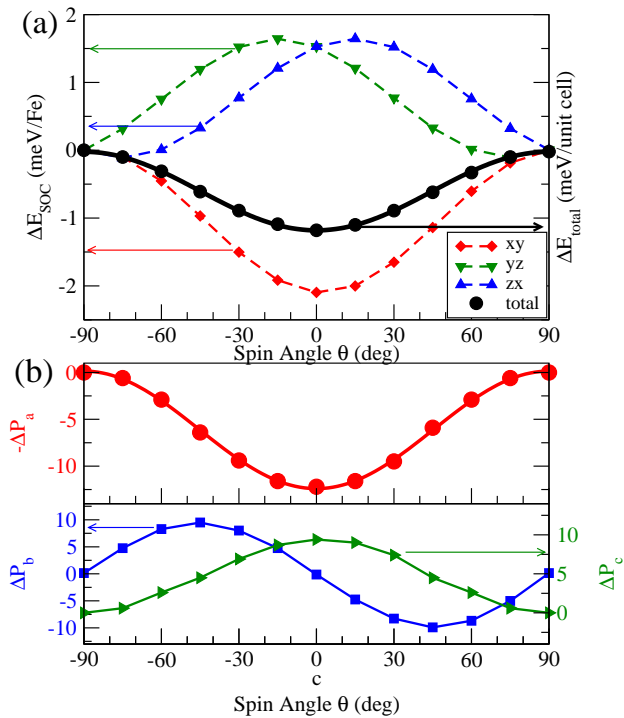}}
%\vspace{1cm}
%\\
%\resizebox{82mm}{!}
%{\includegraphics{figure/mae.ab.eps}}
\caption{\label{fig:DFT} 
DFT results: 
(a) 
%\color{red}
Change in 
%\color{black}
on-site SOC energy ($E_{\rm SOC}$)  and total energy ($E_{\rm total}$) versus collinear spin direction 
%\color{red}
($\theta$)
%\color{black}
 in the $bc$ plane 
%\color{red}
with respect to the $b$ axis. 
%\color{black}
%The label $xy$/$yz$/$zx$ denotes the SOC energy originating from each representative $xy$/$yz$/$zx$ Fe$^{2+}$-B site. 
%\color{red}
The label $xy$/$yz$/$zx$ denotes the each representative Fe$^{2+}$-B site. 
%\color{black}
%The energy difference  is taken with respect to the $b$ axis. 
(b) Change in $\bm P$ ($\mu$C/m$^{-2}$) versus 
%collinear spin direction in the $bc$ plane, 
%\color{red}
$\theta$
%\color{black}
with respect to $P_{a}$=-3.86, $P_{b}$=0 and $P_{c}$=4.89$ \mu$C/cm$^{-2}$ at $\bm M$//$\bm b$.  
In order to compare with experimental results (Ref. \onlinecite{rado.prb1975}), the sign of $\Delta P_{a}$ is reversed and 
 fitted to a function of $a_{1}\sin^{2}\theta + a_{2}\sin^{4}\theta$, with $a_{1}$=1.24$\times$10$^{-3}$ and $a_{2}$=1.29$\times$10$^{-5}$ (red solid line).
}
\vspace{-0.6cm}
\end{figure}
%
%{\em Orbital ordering and Magnetic anisotropy}
%As previously reported,\cite{yamauchi.prb} both the $P2/c$ and $Cc$ structures show CO and orbital ordering (OO) at Fe-B sites. 
%In the unit cell, 32 Fe-B-sites are split into 16 Fe$^{3+}$ ($d^{5}$) sites and 16 Fe$^{2+}$ ($d^{6}$) sites. 
%The latter are Jahn-Teller (JT) active ions, where the partially filled down-spin $t_{2g}^{1\downarrow}$ shell shows one of $d_{xy}$/$d_{yz}$/$d_{zx}$ charge distribution, according to the related JT distortion of the FeO$_{6}$ octahedral cage. Majority spin states, on the other hand, are completely filled. 
%In this context, the JT-acrive 16 Fe$^{2+}$ sites are furthermore split into three groups: 8 with $d_{xy}$, 4 with $d_{yz}$ and 4 with $d_{zx}$ OO. 
%The OO pattern (in this paper, the suffixes $x$, $y$, $z$ are always consistent with the vectors shown in Fig.\ref{fig:parchg}) is such that orbitals avoid to overlap, consistent with reducing the inter-site Coulomb repulsion and optimizing cooperative JT distortions. 
%In this respect, we also note that the FeO6 octahedra are slightly tilted, so that these orbitals do not fully belong to $zx$/$xy$/$zy$ planes.

As for the MAE, our results are consistent with experimental observations, i.e. of
the magnetically hard $\bm a$, intermediate $\bm b$ and easy $\bm c$ axes\cite{rado.prb1975}  
(with  energy differences,  $E_{a}-E_{b}$ = 4.2 meV/cell and  $E_{b}-E_{c}$ = 1.2 meV/cell). 
Experimentally, the large $ab$-plane magnetic anisotropy is well known and used for the
$H$-field cooling technique,  labeled ``magnetic annealing'', which is generally performed to grow large single-domain crystals below $T_{\rm V}$.\cite{rado.prb1975}  
Following the procedure of Ref. \onlinecite{yamauchi.linipo4}, the ``global'' MAE can be decomposed into the ``local'' MAE, 
evaluated by the on-site SOC energy, $E_{\rm SOC}$ \cite{Esoc} at each site. 
Figure \ref{fig:DFT} shows the ``global'' MAE and some selected ``local'' MAEs, upon rotation of Fe spins in the $ab$ plane. 
Each local MAE shows a $\cos2\theta$ curve, which can be explained by SOC perturbation theory. 
On Fe$^{3+}$ ($d^{5}$) ions, a large MAE is not expected, due to the quenched orbital states; on the other hand,  
on  Fe$^{2+}$ ($d^{6}$) ions, we can apply the same discussion used for the Co$^{3+}$($d^{6}$) ion in BiCoO$_{3}$\cite{uratani}. 
%the $d^{6}$ situation,  with the highest occupied $d_{xy}$ state, by analogy with Co$^{3+}$($d^{6}$) case in BiCoO$_{3}$\cite{uratani}, 
The perturbation theory predicts that, when the $d_{xy}^{\downarrow}$ orbital is occupied, 
the single site MAE is given by $E_{\rm MAE}\simeq \lambda^{2}/2(1/\Delta_{\rm JT}-4/\Delta_{e_{g}})\cos2\theta$, 
where $\lambda$ is the SOC constant, $\Delta_{\rm JT}$ is the JT splitting energy, $\Delta_{e_{g}}$ is the energy splitting between $(yz,zx)^{\downarrow}$ and $e_{g}^{\uparrow}$ states, and $\theta$ is the spin polar angle, giving energy minimum at $\theta$=0 ($\bm S$//$\bm z$). 
Our results therefore show that the local magnetic easy axis is perpendicular to the occupied orbital plane (the latter being a hard plane); in other words, the local easy axis  at ${xy}$/${yz}$/${zx}$ site is the $z$/$x$/$y$ axis, respectively. 
Indeed, in Fig. \ref{fig:DFT}, the $xy$-labeled MAE comes from the ${xy}$ site, showing $\cos2\theta$ curve with the minimum at $\bm M//\bm z$($\simeq\bm c$), whereas MAE from ${yz}$ and ${zx}$ site shows maximum nearby.  
The ``global'' MAE is the composition of these local MAEs, based on a delicate balance of (8${xy}$, 4${yz}$, 4${zx}$) OO set; the subtle deviation of the local easy axis away from lattice vectors, due to the small tilting of the orbitals, also affects the final result. 

As for the focus of the paper, {\em i.e.} ME effects, we indeed obtained  $\bm P^{\rm ME}(\theta)$ (cfr Fig.\ref{fig:DFT}(b)),   
showing excellent agreement both with what expected from Landau theory in Eq.\ref{eq:P2} and with experimental results. Unfortunately, the experimental size of $P^{\rm ME}$ has never been reported so far, so that our comparison cannot be quantitative. In our simulations, 
we got $P^{\rm ME}\lesssim 20$$\mu$C/m$^{-2}$. This is roughly 10$^{3}$ times smaller than 
$P^{\rm CO}\simeq 5$$\mu$C/cm$^{-2}$ 
 and even one order of magnitude smaller than $P^{\rm ME}$=800$\mu$C/m$^{-2}$ in a representative ME system, TbMnO$_{3}$ \cite{kimura.tbmno3}; 
 however, it should be large enough to be measured, 
 comparable with observed $P^{\rm ME}$=5$\mu$C/m$^{-2}$ in RbFe(MoO$_{4}$)$_{2}$.\cite{kenzelmann.rfmo} 
Notably, 
%\color{black}
 the $P_b$ trend put forward an exciting perspective: its sign can be switched by a 90$^o$-rotation of the magnetic field (from -45$^o$ to 45$^o$ in Fig. \ref{fig:DFT}(b)), pointing to a magnetic control of ferroelectric polarization.\cite{sym.Pb}

%Before we discuss magnetoelectricity, let us examine how Fe-$d$ orbital states are affected by the  spin rotation  in the $bc$ plane. 
%
In order to understand the mechanism driving ME effects, let us examine how Fe-$d$ orbital states are affected by the  spin rotation  in the $bc$ plane.
\begin{table}[h] 
\vspace{-0.5cm}
\begin{center}
\caption{$3d$-orbital coefficients (in percentage, with spin states summed up) of the highest occupied $d$ orbital and Fe-$d$ occupancy $n_{d}$ on Fe$^{2+}$ B12 site with different SOC enhancement factors $\lambda$ (0=without SOC, $\times$1=with standard SOC, and $\times$10=with the  SOC term 10 times enhanced) for different $\bm M$ directions. } %%%%%In the Berry-phase approach, the values are relative to %%%%%%%the value $U$ = 8 eV (see below).}
\label{tbl.coeff}
\begin{tabular}{|cc|ccccc|c|}
\hline %\hline
%%%%%&$P_{th}$ &\\
%%%%%\hline
%Fe24 occupancy
$\lambda$	&$\bm M$		&$xy$ & ${yz}$ & ${zx}$& $3z^{2}$-$r^{2}$& $x^{2}$-$y^{2}$ & $n_{\rm d}$ \\
\hline
0			&-		&98.85 & 0.02 & 0.37 & 0.76 & 0 & 6.097  \\
\hline
$\times$1			&$\bm M$//$\bm b$	&97.97 & 0.4 & 0.45 & 0.78 & 0.41& 6.083 \\
$\times$1			&$\bm M$//$\bm c$	&98.25 & 0.4 & 0.42 & 0.82 & 0.1 & 6.083\\	   
\hline 
$\times$10	&$\bm M$//$\bm b$	&\bf 51.74 &\bf 17.25 &\bf 16.82 & 1.94 & 12.24 &\bf 6.108\\
$\times$10	&$\bm M$//$\bm c$	&\bf 73.82 &\bf 7.92 &\bf 7.7 & 0.1 & 10.46 &\bf 6.097\\
\hline
\end{tabular}
\end{center}
%to save space 
\vspace{-0.6cm}
\end{table}
Table \ref{tbl.coeff} summarizes the coefficients of a
linear combination of $d$-orbital states for the highest occupied $d$ level \cite{calc.orbital} and the Fe-$d$ occupancy at Fe-B12($xy$) site when $\bm M$//$\bm b$ and $\bm c$.
%, which is relevant to the $\bm P^{\rm CO}$ due to the lack of the inversion counterpart, so that it is probably related to the $\bm P^{\rm PE}$, too,  
% with $\bm M$//$\bm b$ and $\bm c$. 
To highlight the effects of SOC, three calculations were done (keeping the ions fixed): (i) without SOC term ({\em i.e.} where the spin direction doesn't affect the orbitals), (ii) with standard SOC term, and (iii) with  SOC artificially
enhanced by a factor of 10. 
At first glance, $d_{xy}$ shows the largest component in all cases. 
In the extreme case with 10-times-enhanced SOC, 
$d_{yz}$ and $d_{zx}$ orbitals have relatively large components, increasing to about 17\%\ when $\bm M$//$\bm b$, compared to only $\sim$ 8\%\ when $\bm M$//$\bm c$.
This result reflects the ``inverse effect'' of the perturbation theory discussed above, $i.e.$
{\em the orbital plane of the partly filled minority-spin state
 tends to be perpendicular to the imposed spin direction}, despite the fact that the structural JT distortion would favor a $d_{xy}$ state. 
%In this situation, the enhanced SOC term messes up CEF splitting and/or exchange splitting. 
%Focusing of the second highest occupied state, which was originally $xy$ without SOC, 
%the coefficient of $xy$/$yz$/$zx$ changes according to direction of spin, where orbital favours perpendicular to the spin direction, showing agreement with predicted by the perturbation theory.  
The SOC also changes the occupancy of $d$ electrons according to the spin direction. Although the change is negligibly small in standard SOC calculation, it is probably responsible for the slight change in $\bm P^{\rm ME}(\theta)$. 
%Tab.\ref{tbl.coeff} shows that the occupancy of the local Fe-B12 site increases when the spin direction matches the local easy axis. 
%This anticipates an important concept that will be expanded below: 
%{\em since ferroelectric polarization arises from the electric dipole connecting B12 and B14 Fe sites, the modulation of electric charge at one site can change the size of the dipole and the consequent electric polarization}. \\
%In case of ``standard" (non-enhanced) SOC term, the dependence of the occupancy on the spin rotation is not visible within our numerical accuracy. 
Considering the above evidences, we propose the following scenario for magnetoelectricity: 
i) SOC changes the shape of $t_{2g}^{1\downarrow}$ orbital when $\bm M$ is rotated (as shown in Fig.\ref{fig:parchg}(b)). 
ii) this in turn results in the anisotropic hybridization of the Fe-$d$ state with surrounding O-$p$ states, so that the gravity center of the $d$-electron ({\em i.e.} the Wannier center) is slightly shifted in one direction. iii) On the B12 site, the effect doesn't cancel out due to the lack of inversion-related ion in the unit cell, so that a purely electronic contribution appears in the net $\bm P^{\rm orb}$. In other words, the fact that the underlying $Cc$ symmetry lacks inversion is a necessary ingredient for ME effects to emerge. %Note also that the induced orbital moment is not particularly large ($|L|$=0.08$\mu_{B}$, whereas $|S|$=2.6$\mu_{B}$), with  $\bm L$  parallel to $\bm M$, as often seen in more-than-half-filled electron systems.)

%{\em Conclusion.} --- 
 
%We have shown that the origin of the ME effect in magnetite is ``orbital-related" and $\bm P^{\rm orb}$ is induced by the on-site SOC term. 
%This is clearly different from common conventional mechanisms (reviewed in Ref. \onlinecite{fiebig.ME.review}), which invoke inter-site spin interaction, such as magneto-striction and DM interaction. %{\bf HIGHLIGHT THE DIFFERENCE WITH RESPECT TO ON-SITE ME}

{\em Conclusion.} --- We have shown that the origin of the ME effect in magnetite is ``orbital-related" and $\bm P^{\rm orb}$ is induced by the on-site SOC term. 
This is clearly different from common conventional mechanisms (reviewed in Ref.\cite{fiebig.ME.review}), which invoke inter-site spin interaction, such as magneto-striction and DM interaction, in turn often connected to non-collinear spin-configurations. Rather, magnetoelectricity in magnetite emerge in a fully-collinear spin state with large magnetization, pointing to an easier control of ferroelectric properties via magnetic fields.%{\bf HIGHLIGHT THE DIFFERENCE WITH RESPECT TO ON-SITE ME} 
We also note that the on-site $\bm P^{\rm orb}$ is not coupled with ``structural" effects, such as piezo-electric or piezo-magnetic. 
%In addition, mechanism of single-site anisotropy regarded as the origin of ME effect at Cr$_{2}$O$_{3}$
% is also different since it requires the ionic displacement according to the phonon mode and the spin-canting under $\bm H$.\cite{cr2o3.1, cr2o3.2} 
Rather, the main ingredient which allows magnetite to be ME-active is basically the polar OO pattern of unfilled $t_{2g}$ states. Indeed, no ionic displacements are involved in the rise of H--induced polarization and, in this sense, we can label this novel ME mechanism as {\em purely driven by electronic (charge and orbital) degrees of freedom}. Remarkably, our predicted trends show that the long-sought full control of polarization via a magnetic field can be achieved, with $P_b$ switching sign upon a 90$^o$ rotation of H-field. 
%It comes into play in magnetite for several reasons: 
We finally recall that the polar OO is stabilized together with the polar CO in the $Cc$ structure,  
through the CDW instability coupled with the Fermi-surface nesting below $T_{\rm V}$.\cite{wright.prl2001} %, yanase}
%for several reasons; 
%the softening phonon mode coupled with the CDW instability, 
%(or Peierls instability) 
%which enables the Fermi-surface nesting,  enlarges the crystal cell below $T_{\rm V}$,  
% the $Cc$ symmetry, 
%as allowing for the polar CO/OO pattern.    
These peculiar properties make magnetite rather unique; however, we propose at least one candidate where the phenomenology might be similar:  K$_{0.6}$FeF$_3$,\cite{yamauchi.ttb} showing a non-centrosymmetric CO/OO pattern and where ME effects should be explored, to confirm the existence of a novel class of OO-induced magnetoelectrics.

\acknowledgments
KY thanks A. Tanaka, T. Shishidou, T. Oguchi, T. Kimura, C. Ederer and M. Angst for fruitful discussions.
The research leading to these results has received funding from the EU Seventh Framework Programme  (FP7/2007-2013) under the ERC grant agreement n. 203523-BISMUTH.
Computational support from Caspur  Supercomputing Center  (Rome) is also acknowledged. 

\begin{thebibliography}{200}
\bibitem{lufeo} LuFe$_{2}$O$_{4}$ was originally proposed as a prototype of this class of ferroelectrics\protect{\cite{ikeda}}, but turned out to be anti-ferroelectric\protect{\cite{angst}} later on.
\bibitem{suffix} In this paper, the suffixes of $d_{xy}$, $d_{yz}$ and $d_{zx}$ are always consistent with the ($x, y, z$) vectors shown in Fig.\ref{fig:parchg}. We denote the corresponding Fe$^{2+}$ site as simply $xy$, $yz$ and $zx$ site, respectively.
\bibitem{Esoc}
The SOC energy, $E_{\rm SOC}=\langle \frac{1}{c^{2}} \frac{1}{r} \frac{dV}{dr} l\cdot s\rangle$ is integrated in each atomic sphere, where $c$ is the velocity of light, $r$ is the radial distance in each atomic sphere, $V$ is the effective potential as a function of $r$, and  $l$ and $s$ are respectively orbital and spin operators. Also see Ref. \onlinecite{yamauchi.linipo4}.
\bibitem{sym.Pb}
From the symmetry point of view, $P_{b}$ is not allowed, due to the $c$-glide in the $Cc$ space group. 
When considering the magnetic space groups, the $c$-glide is kept as $c$ ($Tc$; with time reversal) when $\bm M$ is parallel to $\bm b$ (to $\bm c$). 
However, when $\bm M$ is rotated between $\bm b$ and $\bm c$, no symmetry exists (except for the identity $E$), so that any direction of $\bm P$ is allowed, including $P_{b}$ which is purely SOC induced.
\bibitem{calc.orbital} 
The coefficients of $d$-orbitals are obtained by diagonalizing an effective on-site Hamiltonian: 
$H_{ij}=1/2\delta_{ij}-\left|i\right>\left<j\right| U_{\rm eff}$, where $\left|i\right>\left<j\right|$ is 10$\times$10 density matrix (5 orbitals $\times$ 2 spins) used in GGA+$U$ procedure and $U_{\rm eff}$ is an effective on-site Coulomb parameter, described as $U_{\rm eff}$=$U$-$J$=3.61eV by GGA+$U$ parameters. The occupancy is obtained by the trace of density matrix.
\end{thebibliography}
\end{document}